\documentstyle[preprint,prd,aps]{revtex}

\tightenlines

\begin{document}
\title{{\bf Modified initial state for perturbative QCD}}

\author{{\bf Marcos Rigol Madrazo} \\ {\it Centro de Estudios Aplicados al
Desarrollo Nuclear} \\ {\it Calle 30, N. 502 e/ 5ta y 7ma,
Miramar, La Habana, Cuba } \\ \vspace{0.5cm} {\bf Alejandro Cabo
Montes de Oca} \\ {\it Abdus Salam International Centre for
Theoretical Physics} \\ {\it Strada Costiera 11, 34014, Trieste,
Italy} \\ {\it and } \\ {\it Instituto de Cibern\'{e}tica
Matem\'{a}tica y F\'{\i}sica} \\ {\it Calle E, N. 309, Esq. a 15,
Vedado, La Habana, Cuba} \\ \vspace{1cm} (Published in Phys. Rev. D62, 
074018, 2000)
\\ \vspace{1cm} (Final Version)}
\date{September, 2001}

\maketitle

\vspace{1cm}

\begin{abstract}
A particular initial state for the construction of the
perturbative expansion of QCD is investigated. It is formed as a
coherent superposition of zero momentum gluon pairs and shows
Lorentz as well as global $SU(3)$ symmetries. It follows that the
gluon and ghost propagators determined by it, coincides with the
ones used in an alternative of the usual perturbation theory
proposed in a previous work. Therefore, the ability of such a
procedure of producing a finite gluon condensation parameter
already in the first orders of perturbation theory is naturally
explained. It also follows that this state satisfies the
physicality condition of the BRST procedure in its Kugo and Ojima
formulation. The BRST quantization is done for the value
$\alpha=1$ of the gauge parameter where the procedure is greatly
simplified. Therefore, after assuming that the adiabatic
connection of the interaction does not take out the state from the
interacting physical space, the predictions of the perturbation
expansion, at the value $ \alpha=1$, for the physical quantities
should have meaning. The validity of this conclusion solves the
gauge dependence indeterminacy remained in the proposed
perturbation expansion.
\end{abstract}

\newpage

\section{Introduction.}

Quantum Chromodynamics (QCD) was discovered in the seventies and
up to this time it is considered as the fundamental theory for the
strong interactions, as a consequence it has been deeply
investigated \cite{Yang}.

In one limit, the smallness of the coupling constant at high
momentum (asymptotic freedom) made possible the theoretical
investigation of the so-called hard processes by using the
familiar perturbative language. This so-called Perturbative QCD
(PQCD) was satisfactorily developed. However, relevant phenomena
associated with the strong interactions can't be described by the
standard perturbative methods and the development of the
Non-Perturbative QCD is at the moment one of the challenges of
this theory.

One of the most peculiar characteristics of the strong
interactions is color confinement. According to this philosophy,
colored objects, like quarks and gluons, can't be observed as free
particles in contrast with hadrons that are colorless composite
states and effectively detected. The physical nature of such
phenomenon remains unclear. Numerous attempts to explain this
property have been made, for example explicit calculations in
which the theory is regularized on a spatial lattice
\cite{Creutz}, and also through the construction of
phenomenological models. In the so called MIT Bag Model
\cite{Chodos}, it is assumed that a bag or a bubble is formed
around the objects having color in such a way that they could not
escape from it, because their effective mass is smaller inside the
bag volume and very high outside. Within the so called String
Model \cite{Gervais}, which is based in the assumption that the
interaction forces between quarks and antiquarks grow when the
distance increases, in such a way that the energy increases
linearly with the string length $E(L)=kL$.

A fundamental problem in QCD is the nature of the ground state
\cite{Shuryak1,ShuryakTex,Shuryak2}. This state is imagined as a
very dense state of matter, composed of gluons and quarks
interacting in a complicated way. Its properties are not easily
accessible in experiments, because quarks and gluons fields cannot
be directly observed. Furthermore, the interactions between quarks
can't be directly determined.

It is already accepted that in QCD the zero point oscillations of
the coupled modes produce a finite energy density, which is
determined phenomenologically. The numerical estimate of it is

\[
E_{vac} \simeq -f \langle 0\mid (gG)^2\mid 0\rangle \simeq
0.5GeV/fm^3,
\]
\noindent where the so called non perturbative gluonic condensate
$\langle 0\mid(gG)^2 \mid 0\rangle $ was introduced and
phenomenologically evaluated by Shifman, Vainshtein and Zakharov
\cite{Zakharov}. The negative sign of $E_{vac}$ means that the non
perturbative vacuum energy is lower that the one associated to the
perturbative vacuum.

Since a long time ago, one particular kind of models has shown to
be able to predict similar properties. These are the
chromo-magnetic vacuum approaches, in which it is assumed that a
vacuum magnetic field is existing at all the points \cite{Savv2}.
Concretely a constant magnetic abelian field $H$ is assumed. A
one-loop calculation gives as the result the following energy
density

\[
E\left(H\right) =\frac{H^2}2\left(1+\frac{bg^2}{16\pi ^2} \ln
\left(\frac H{\Lambda ^2}\right) \right).
\]

This formula predicts negative values for the energy at small
fields, so the usual perturbative ground state with $H=0$ is
unstable with respect to the formation of a state with a non
vanishing field intensity at which the energy $E\left(H\right)$
has a minimum \cite{Savv2}. Many physical problems related with
hadron structure, confinement problem, etc. have been investigated
using the Savvidy model. Nevertheless, after some time its intense
study was abandoned. The main reasons were: (1) The perturbative
relation giving $E_{vac}$ would only be valid if the second order
in the expansion in powers of the Planck constant is relatively
small. (2) The specific spatial direction and the color direction
of the magnetic field break the now seemingly indispensable
Lorentz and $SU(3)$ invariance of the ground state. (3) The
magnetic moment of the vector particle (gluon) is such that its
energy in the presence of the field has a negative eigenvalue,
which also makes the homogeneous magnetic field $H$ unstable.

Before presenting the objectives of the present work it should be
stressed that the perturbative quantization of QCD is realized in
the same way as that in QED. The quadratic field terms of the QCD
Lagrangian have the same form that the ones corresponding to the
electrons and photons in QED. However, in connection with the
interaction, there appear a substantial difference due to the
coupling of the gluon to itself. In addition is a general fact
that a perturbative expansion has some freedom in dependence on
the initial conditions at $t\rightarrow \pm \infty $ or what is
the same, from the states in which the expansion is based.
Moreover, as was expressed before, the exact ground state has a
non-trivial structure associated to a gluon condensate.

Then, given the above remarks, it is not unreasonable to expect
that the true vacuum state could be well described through a
modified Feynman expansion perturbatively describing a gluon
condensate. Such a perturbative condensate could generate all the
low energy physical observable, which in the standard expansion
could require an infinite number of terms of the series.

In a previous work of one of the authors (A.C.) \cite{Cabo},
following the above idea, a modified perturbation theory for QCD
was proposed. This expansion retains the main invariance of the
theory (the Lorentz and $SU(3)$ ones), and is also able to
reproduce main physical predictions of the chromomagnetic field
models. It seems possible to us that this procedure could produce
a reasonable if not good description of the low energy physics. If
it is the case, then, the low and high energy descriptions of QCD
would be unified in a common perturbative framework. In particular
in \cite{Cabo} the results had the interesting outcome of
producing a non-vanishing mean value for the relevant quantity
$\langle G^2\rangle $. In addition the effective potential for the
condensation parameter in the first order approximation shows a
minimum at non vanishing values of that parameter. Therefore, the
procedure is able to reproduce at least some central predictions
of the chromomagnetic models and general QCD analysis.

The main objective of the present work consists in investigating
the foundations of the mentioned perturbation theory. The concrete
aim is to find a state in the Fock space of the non interacting
theory being able to generate that expansion by also satisfying
the physicality condition of the BRST quantization approach.

It follows that it is possible to find the sought for state and it
turns out to be an exponential of a product of two gluon and ghost
creation operators. That is, it can be interpreted as a coherent
superposition of states with many zero momentum gluon and ghost
pairs. Therefore, this structure gives an explanation for the
ability of the expansion being investigated to produce non zero
values of the gluon condensation in the first orders of
perturbation theory. The fact that the effective action also shows
a minimum for non-vanishing values of the condensation parameter
also supports the idea that the considered state improves the
perturbative expansion. It is also shown that the state satisfies
the linear condition which defines the physical subspace in the
BRST quantization for the $\alpha =1$ value of the gauge
parameter. Thus, the indefiniteness in the appropriate value of
this parameter to be used which remained in the former work is
resolved opening the way for the study of the predictions of the
proposed expansion.

It should mentioned that a similar idea as the one advanced in
\cite{Cabo} was afterwards considered in \cite{hoyer}. In that
work, gluon and quark condensation in a range of momenta of the
order of $\Lambda _{qcd}$ have been considered with the similar
aim of constructing an alternative perturbative theory for QCD.
However, in our view, that construction should break the Lorentz
invariance because the condensates are expected to show a
non-vanishing energy density. If this limitation is not at work
their proposed mechanism could be worth considering as an
alternative possibility.

The exposition will be organized as follows: In Section 2, the
BRST operational quantization method for gauge fields developed by
Kugo and Ojima is reviewed. Starting from it, in Section 3 the
ansatz for the Fock space state, which generates the desired form
of the perturbative expansion, is introduced. The proof that the
state satisfies the physical state condition is also given in this
section. Then, in Section 4 it is shown that the proposed state
can generate the wanted modification of the propagator by a proper
selection of the parameters at hand. The modification of the
propagator for the ghost particles is also considered in this
section. Finally the evaluation of the gluon condensation
parameter done in a previous work \cite{Cabo} is reviewed in order
to illustrate the ability of the procedure in predicting a main
property of the real QCD ground state.

\section{Review of the K-O Quantization procedure}

In the present section the operator formalism developed by T. Kugo
and I. Ojima \cite{Kugo1,Kugo2,Kugo3,Kugo4} is reviewed and after
specialized to the non interacting limit of gluodynamics (GD).
This formulation takes into account the invariance of the
Lagrangian under a global symmetry operation called the BRST
transformation \cite{BRST}. We will consider the construction of a
relativistic invariant initial state in the non-interacting limit
of QCD. The BRST physical state condition in the non-interacting
limit will be also imposed. As explained before, the motivation is
that we think that this state has the possibility to furnish the
gluodynamics ground state under the adiabatic connection of the
interaction. In what follows we will work in Minkowski space with
the conventions defined below.

Let G a compact group and $\Lambda$ any matrix in the adjoint
representation of its associated Lie Algebra. The matrix $\Lambda$
can be represented as a linear combination of the form

\begin{equation}
\Lambda =\Lambda ^aT^a,
\end{equation}

\noindent where $T^a$ are the generators $(a=1,...,$Dim$ G=n)$
which are chosen as Hermitian ones, satisfying

\begin{equation}
\left[ T^a,T^b\right] =if^{abc}T^c.
\end{equation}

The variation of the fields under infinitesimal gauge
transformations are given by

\begin{eqnarray}
\delta _\Lambda A_\mu ^a\left(x\right) &=&\partial _\mu \Lambda
^a\left(x\right) +gf^{acb}A_\mu ^c\left(x\right) \Lambda
^b\left(x\right) =D_\mu ^{ab}\left(x\right) \Lambda ^b, \\ D_\mu
^{ab}\left(x\right) &=&\partial _\mu \delta ^{ab}+gf^{acb}A_\mu
^c\left(x\right).
\end{eqnarray}

The metric $g_{\mu \nu }$ is diagonal and taken in the convention

\begin{equation}
g_{00}=-g_{ii}=1 \qquad \text{for}\ \ \, i=1,2,3.
\end{equation}

The complete GD Lagrangian to be considered is the one employed in
the operator quantization approach of \cite{OjimaTex}. Its
explicit form is given by

\begin{eqnarray}
{\cal L} &=&{\cal L}_{YM}+{\cal L}_{GF}+{\cal L}_{FP}  \label{Lag}
\\ {\cal L}_{YM} &=&-\frac 14F_{\mu \nu }^a\left(x\right) F^{\mu
\nu,a}\left(x\right),  \label{YM} \\ {\cal L}_{GF} &=&-\partial
^\mu B^a\left(x\right) A_\mu ^a\left(x\right) + \frac \alpha
2B^a\left(x\right) B^a\left(x\right),  \label{GF} \\ {\cal L}_{FP}
&=&-i\partial ^\mu \overline{c}^a\left(x\right) D_\mu
^{ab}\left(x\right) c^b\left(x\right).  \label{FP}
\end{eqnarray}

The field strength is

\begin{equation}
F_{\mu \nu }^a\left(x\right) =\partial _\mu A_\nu ^a\left(x\right)
-\partial _\nu A_\mu ^a\left(x\right) +gf^{abc}A_\mu
^b\left(x\right) A_\nu ^c\left(x\right) .
\end{equation}

Relation (\ref{YM}) defines the Yang-Mills standard Lagrangian,
(\ref{GF}) is the gauge fixing term and (\ref{FP}) is the
Lagrangian which describes the dynamics of the nonphysical
Faddeev-Popov ghost fields.

The physical state conditions in the BRST procedure
\cite{OjimaTex,Govaerts} are given by

\begin{eqnarray}
&&Q_B\mid \Phi \rangle =0, \\
&&Q_C\mid \Phi \rangle =0.
\end{eqnarray}

\noindent where

\begin{eqnarray}
&&Q_B=\int d^3x\left[ B^a\left(x\right) \nabla _0
c^a\left(x\right) +gB^a\left(x\right) f^{abc}A_0^b\left(x\right)
c^c\left(x\right) \right. \nonumber \\ &&\hspace{6cm}\left. +\frac
i2g\partial _0\left(\overline{c}^a\right) f^{abc}c^b\left(x\right)
c^c\left(x\right) \right],
\end{eqnarray}

\noindent with

\begin{equation}
f\left(x\right) \nabla_0 g\left(x\right) \equiv f\left(x\right)
\partial _0g\left(x\right) -\partial _0\left(f\left(x\right)
\right) g\left(x\right).
\end{equation}

The BRST charge is conserved as a consequence of the BRST symmetry
of Lagrangian (\ref{Lag}). The also conserved charge $Q_C$ is
given by

\begin{equation}
Q_C=i\int d^3x\left[ \overline{c}^a\left(x\right) \nabla_0 c^a
\left(x\right) +g\overline{c}^a\left(x\right)
f^{abc}A_0^b\left(x\right) c^c\left(x\right) \right],
\end{equation}

\noindent which conservation comes from the Noether theorem due to
the invariance of Lagrangian (\ref{Lag}) under the phase
transformation $ c\rightarrow e^\theta c,\ \overline{c}\rightarrow
e^{-\theta }\overline{c}$. This charge defines the so-called
``ghost number'' as the difference between the number of $c$ and
$\overline{c}$ ghosts.

Our interest will be centered here in the Yang-Mills theory
without spontaneous breaking of the gauge symmetry in the limit of
no interaction. The quantization of the theory defined by the
Lagrangian (\ref{Lag}), after to be considered in the interaction
free limit $g\rightarrow 0$, leads to the following commutation
relations for the free fields

\begin{eqnarray}
\left[ A_\mu ^a\left(x\right),A_\nu ^b\left(y\right) \right]
&=&\delta ^{ab}\left(-ig_{\mu \nu }D\left(x-y\right)
+i\left(1-\alpha \right)
\partial _\mu \partial _\nu E\left(x-y\right) \right),  \nonumber \\
\left[ A_\mu ^a\left(x\right),B^b\left(y\right) \right] &=&\delta
^{ab}\left(-i\partial _\mu D\left(x-y\right) \right), \nonumber
\\ \left[ B^a\left(x\right) ,B^b\left(y\right) \right]
&=&\left\{ \overline{c}
^a\left(x\right),\overline{c}^b\left(y\right) \right\} =\left\{
c^a\left(x\right),c^b\left(y\right) \right\} =0,  \nonumber \\
\left\{ c^a\left(x\right),\overline{c}^b\left(y\right) \right\}
&=&-D\left(x-y\right).  \label{com}
\end{eqnarray}

The equations of motion for the non-interacting fields takes the
simple form

\begin{equation}
\Box A_\mu ^a\left(x\right) -\left(1-\alpha \right) \partial _\mu
B^a\left(x\right) =0,
\end{equation}
\begin{equation}
\partial ^\mu A_\mu ^a\left(x\right) +\alpha B^a\left(x\right)
=0,
\label{liga1}
\end{equation}
\begin{equation}
\Box B^a\left(x\right) =\Box c^a\left(x\right) =\Box
\overline{c}^a\left(x\right) =0.
\end{equation}

This equations can be solved for arbitrary values of $\alpha$.
However, as it was said before, the discussion will be restricted
to the case $\alpha=1$ which corresponds to the situation in which
all the gluon components satisfy the D'Alambert equation. This
selection, as considered in the framework of the usual
perturbative expansion, implies that you are not able to check the
$\alpha $ independence of the physical quantities. However in the
present discussion the aim is to construct a perturbative state
that satisfies the BRST physicality condition, in order to connect
adiabatically the interaction. Then, the physical character of all
the prediction will follow whenever the former assumption that
adiabatic connection does not take the state out of the physical
subspace at any intermediate state is valid. Clearly, the
consideration of different values of $\alpha $, would be also a
convenient recourse for checking the $\alpha \ $independent
perturbative expansion. However, at this stage it is preferred to
delay the more technical issue of implementing the BRST
quantization for any value of $ \alpha$ for future work.

In that way the field equations in the $\alpha =1$ gauge will be

\begin{equation}
\Box A_\mu ^a\left(x\right) =0,  \label{movi1}
\end{equation}
\begin{equation}
\partial ^\mu A_\mu ^a\left(x\right)+ B^a\left(x\right) =0, \label{movi3}
\end{equation}
\begin{equation}
\Box B^a\left(x\right) =\Box c^a\left(x\right) =\Box
\overline{c}^a\left(x\right) =0.  \label{movi2}
\end{equation}

The solutions of the set (\ref{movi1})-(\ref{movi2}) can be
written as

\begin{eqnarray}
A_\mu ^a\left(x\right) &=&\sum\limits_{\vec{k},\sigma
}\left(A_{\vec{k} ,\sigma }^af_{k,\mu }^\sigma \left(x\right)
+A_{\vec{k},\sigma }^{a+}f_{k,\mu }^{\sigma *}\left(x\right)
\right),  \nonumber \\ B^a\left(x\right)
&=&\sum\limits_{\vec{k}}\left(B_{\vec{k}}^ag_k\left(x\right)
+B_{\vec{k}}^{a+}g_k^{*}\left(x\right) \right),  \nonumber \\
c^a\left(x\right)
&=&\sum\limits_{\vec{k}}\left(c_{\vec{k}}^ag_k\left(x\right)
+c_{\vec{k}}^{a+}g_k^{*}\left(x\right) \right),  \nonumber \\
\overline{c}^a\left(x\right)
&=&\sum\limits_{\vec{k}}\left(\overline{c}_{
\vec{k}}^ag_k\left(x\right)
+\overline{c}_{\vec{k}}^{a+}g_k^{*}\left(x\right) \right).
\end{eqnarray}

\noindent The wave packets for non-massive scalar and vector
fields are taken as

\begin{eqnarray}
g_k\left(x\right) &=&\frac 1{\sqrt{2Vk_0}}\exp \left(-ikx\right) ,
\nonumber \\ f_{k,\sigma }^\mu \left(x\right) &=&\frac
1{\sqrt{2Vk_0}}\epsilon _\sigma ^\mu \left(k\right) \exp
\left(-ikx\right).  \label{pol}
\end{eqnarray}

As can be seen from (\ref{movi3}) the five $A_{\vec{k},\sigma }^a$
and $B_{ \vec{k}}^a\ $ modes are not all independent. Indeed, it
follows from (\ref {movi3}) that

\begin{equation}
B_{\vec{k}}^a=A_{\vec{k}}^{S,a}=A_{\vec{k},L}^a.
\end{equation}

Then, the expansion of the free Heisenberg fields takes the form

\begin{equation}
A_\mu ^a\left(x\right)
=\sum\limits_{\vec{k}}\left(\sum\limits_{\sigma
=1,2}A_{\vec{k},\sigma }^af_{k,\mu }^\sigma \left(x\right)
+A_{\vec{k} }^{L,a}f_{k,L,\mu }\left(x\right)
+B_{\vec{k}}^af_{k,S,\mu }\left(x\right) \right)+h.c.,
\end{equation}

\noindent where $h.c.$ represents the Hermitian conjugate of the
first term. In order to satisfy the commutation relations
(\ref{com}) the creation and annihilation operator associated to
the Fourier components of the fields should obey

\begin{eqnarray}
\left[ A_{\vec{k},\sigma }^a,A_{\vec{k}^{\prime },\sigma ^{\prime
}}^{a^{\prime }+}\right] &=&-\delta ^{aa^{\prime }}\delta
_{\vec{k}\vec{k} ^{\prime }}\tilde{\eta}_{\sigma \sigma ^{\prime
}},  \nonumber \\ \left\{
c_{\vec{k}}^a,\overline{c}_{\vec{k}^{\prime }}^{a^{\prime
}+}\right\} &=&i\delta ^{aa^{\prime }}\delta
_{\vec{k}\vec{k}^{\prime }}, \nonumber \\ \left\{
\overline{c}_{\vec{k}}^a,c_{\vec{k}^{\prime }}^{a^{\prime
}+}\right\} &=&-i\delta ^{aa^{\prime }}\delta
_{\vec{k}\vec{k}^{\prime }}
\end{eqnarray}
and all the others vanish, $\tilde{\eta}_{\sigma \sigma ^{\prime
}}=\epsilon _\sigma ^\mu \left(k\right) \epsilon _{\sigma ^{\prime
},\mu }^{*}\left(k\right) $, for $\sigma,\sigma ^{\prime
}=1,2,L,S$.

The above commutation rules and equations of motion define the
quantized non-interacting limit of GD. It is possible now to start
defining the alternative interaction free ground state to be
considered for the adiabatic connection of the interaction. As
discussed before, the expectation is that the physics of the
perturbation theory being investigated is able to furnish a
helpful description of low energy physical effects.

\section{The alternative initial state}

After beginning to work in the K.O. formalism some indications
were found about that the appropriate state vector obeying the
physical condition in this procedure could have the general
structure

\begin{equation}
|\phi \rangle =\exp \sum\limits_a\left(C_1\left(\left|
\vec{p}\right| \right)
A_{\vec{p},1}^{a+}A_{\vec{p},1}^{a+}+C_2\left(\left|
\vec{p}\right| \right)
A_{\vec{p},2}^{a+}A_{\vec{p},2}^{a+}+C_3\left(\left|
\vec{p}\right| \right)
\left(B_{\vec{p}}^{a+}A_{\vec{p}}^{L,a+}+i\overline{c}_{\vec{p}
}^{a+}c_{\vec{p}}^{a+}\right) \right) \mid 0\rangle \label{Vacuum}
\end{equation}

\noindent where $\vec{p}$ is an auxiliary momentum, chosen as one
of the few smallest values of the spatial momentum for the
quantized theory in a finite volume $V$. This value will be taken
later in the limit $V\rightarrow \infty $ for recovering Lorentz
invariance. From here the sum on the color index $a$ will be
explicit. The parameters $C_i\left(\left| \vec{p}\right| \right)$
will be fixed below from the condition that the free propagator
associated to a state satisfying the BRST physical state
condition, coincides with the one proposed in the previous work
\cite{Cabo}. The solution of this problem, would then give
foundation to the physical implications of the discussion in that
work.

It should also be noticed that the state defined by (\ref{Vacuum})
have some similarity with coherent states \cite{Itzykson}.
However, in the present case, the creation operators appear in
squares. Thus, the argument of the exponential creates pairs of
physical and non-physical particles. An important property of this
function is that its construction in terms of pairs of creation
operators determines that the mean value of an odd number of field
operators vanishes. This is at variance with the standard coherent
state, in that the mean values of the fields are non-zero. The
vanishing of the mean field is a property in common with the
standard perturbative vacuum, which Lorentz invariance could be
broken by any non-zero expectation value of the 4-vector of the
gauge field. It should be also stressed that this state as formed
by the superposition of states of pair of gluons suggests a
connection with some recent works in the literature that consider
the formation of gluons pairs due to color interactions.

Let us argue below that the state (\ref{Vacuum}) satisfies the
BRST physical conditions

\begin{eqnarray}
&&Q_B\mid \Phi \rangle =0, \\
&&Q_C\mid \Phi \rangle =0.
\end{eqnarray}

The expression of the charges in the interaction free limit
\cite{OjimaTex} are

\begin{eqnarray}
Q_B
&=&i\sum\limits_{\vec{k},a}\left(c_{\vec{k}}^{a+}B_{\vec{k}}^a-B_{\vec{
k }}^{a+\ }c_{\vec{k}}^a\right), \\ Q_C
&=&i\sum\limits_{\vec{k},a}\left(\overline{c}_{\vec{k}}^{a+}c_{\vec{k}
}^a+c_{\vec{k}}^{a+}\overline{c}_{\vec{k}}^a\right).
\end{eqnarray}

Considering first the action of $Q_B$

{\setlength\arraycolsep{0.5pt}
\begin{eqnarray}
&&Q_B\mid \Phi \rangle =i\exp \left\{ \sum\limits_{\sigma
,a}C_\sigma \left(\left| \vec{p}\right| \right) A_{\vec{p},\sigma
}^{a+}A_{\vec{p},\sigma }^{a+}\right\} \times  \nonumber \\
&&\times \left(\exp \left\{ \sum\limits_aC_3\left(\left|
\vec{p}\right| \right)
i\overline{c}_{\vec{p}}^{a+}c_{\vec{p}}^{a+}\right\} \sum\limits_{
\vec{k},b}c_{\vec{k}}^{b+}B_{\vec{k}}^b\exp \left\{
\sum\limits_aC_3\left(\left| \vec{p}\right| \right)
B_{\vec{p}}^{a+}A_{\vec{p}}^{L,a+}\right\} \right. \\ &&-\left.
\exp \left\{ \sum\limits_aC_3\left(\left| \vec{p}\right| \right)
B_{\vec{p}}^{a+}A_{\vec{p}}^{L,a+}\right\}
\sum\limits_{\vec{k},b}B_{\vec{k} }^{b+}c_{\vec{k}}^b\exp \left\{
\sum\limits_aC_3\left(\left| \vec{p}\right| \right)
i\overline{c}_{\vec{p}}^{a+}c_{\vec{p}}^{a+}\right\} \right) \mid
0\rangle =0,  \nonumber
\end{eqnarray}
}

\noindent in which the following identity was used

\begin{equation}
\left[ B_{\vec{k}}^b,\exp \sum\limits_aC_3\left(\left|
\vec{p}\right| \right) B_{\vec{p}}^{a+}A_{\vec{p}}^{L,a+}\right]
=-C_3\left(\left| \vec{p} \right| \right) B_{\vec{p}}^{b+}\delta
_{\vec{k},\vec{p}}\exp \sum\limits_aC_3\left(\left| \vec{p}\right|
\right) B_{\vec{p}}^{a+}A_{\vec{ p}}^{L,a+}. \label{ident1}
\end{equation}

For the action of $Q_C$ on the considered state we have

{\setlength\arraycolsep{0.1pt}
\begin{eqnarray}
&&Q_C\mid \Phi \rangle =i\exp \left\{ \sum\limits_{\sigma
,a}C_\sigma \left(\left| \vec{p}\right| \right) A_{\vec{p},\sigma
}^{a+}A_{\vec{p},\sigma }^{a+}+\sum\limits_aC_3\left(\left|
\vec{p}\right| \right) B_{\vec{p} }^{a+}A_{\vec{p}}^{L,a+}\right\}
\\ &&\times \left[ \sum\limits_{\vec{k},b}
\overline{c}_{\vec{k}}^{b+}c_{\vec{k}
}^b\left(1+\sum\limits_aiC_3\left(\left| \vec{p}\right| \right)
\overline{ c }_{\vec{p}}^{a+}c_{\vec{p}}^{a+}\right)
+\sum\limits_{\vec{k},b}c_{\vec{k}
}^{b+}\overline{c}_{\vec{k}}^b\left(1+\sum\limits_aiC_3\left(\left|
\vec{p} \right| \right)
\overline{c}_{\vec{p}}^{a+}c_{\vec{p}}^{a+}\right) \right] \mid
0\rangle =0  \nonumber
\end{eqnarray}
}

\noindent which vanishes due to the commutation rules of the ghost
operators.

Next, the evaluation of the norm of the proposed state is
considered. Due to the commutation properties of the operators, it
can be written as

{\setlength\arraycolsep{0.5pt}
\begin{eqnarray}
\langle \Phi \mid \Phi \rangle =\prod\limits_{a=1,..,8} &\left\{
\prod\limits_{\sigma =1,2}\langle 0\mid \exp \left\{ C_\sigma
^{*}\left(\left| \vec{p}\right| \right) A_{\vec{p},\sigma
}^aA_{\vec{p},\sigma }^a\right\} \exp \left\{ C_\sigma
\left(\left| \vec{p}\right| \right) A_{ \vec{p},\sigma
}^{a+}A_{\vec{p},\sigma }^{a+}\right\} \mid 0\rangle \right. &
\nonumber \\ &\times \langle 0\mid \exp \left\{
C_3^{*}\left(\left| \vec{p}\right| \right)
A_{\vec{p}}^{L,a}B_{\vec{p}}^a\right\} \exp \left\{
C_3\left(\left| \vec{p}\right| \right)
B_{\vec{p}}^{a+}A_{\vec{p}}^{L,a+}\right\} \mid 0\rangle &
\nonumber \\ &\left. \times \langle 0\mid
\left(1-iC_3^{*}\left(\left| \vec{p}\right| \right)
c_{\vec{p}}^a\overline{c}_{\vec{p}}^a\right)
\left(1+iC_3\left(\left| \vec{p}\right| \right)
\overline{c}_{\vec{p}}^{a+}c_{\vec{p} }^{a+}\right) \mid 0\rangle
\right\}.&
\end{eqnarray}
}

For the product of the factors associated to transverse modes and
the eight values of the color index, after expanding the
exponential in series it follows

\begin{eqnarray}
&&\left[ \langle 0\mid \exp \left\{ C_\sigma ^{*}\left(\left|
\vec{p} \right| \right) A_{\vec{p},\sigma }^aA_{\vec{p},\sigma
}^a\right\} \exp \left\{ C_\sigma \left(\left| \vec{p}\right|
\right) A_{\vec{p},\sigma }^{a+}A_{\vec{p},\sigma }^{a+}\right\}
\mid 0\rangle \right] ^8  \nonumber \\ &&=\left[ \langle 0\mid
\sum\limits_{m=0}^\infty \left| C_\sigma \left(\left|
\vec{p}\right| \right) \right| ^{2m}\frac{\left(A_{\vec{p},\sigma
}^a\right) ^{2m}\left(A_{\vec{p},\sigma }^{a+}\right)
^{2m}}{\left(m!\right) ^2}\mid 0\rangle \right] ^8  \nonumber \\
&&=\left[ \sum\limits_{m=0}^\infty \left| C_\sigma \left(\left|
\vec{p} \right| \right) \right| ^{2m}\frac{\left(2m\right)
!}{\left(m!\right) ^2} \right] ^8,  \label{normT}
\end{eqnarray}

\noindent where we used the identity

\begin{equation}
\langle 0\mid \left(A_{\vec{p},\sigma }^a\right)
^{2m}\left(A_{\vec{p} ,\sigma }^{a+}\right) ^{2m}\mid 0\rangle
=\left(2m\right) !.
\end{equation}

The factors related with the scalar and longitudinal modes can be
transformed as follows

\begin{eqnarray}
&&\left[ \langle 0\mid \exp \left\{ C_3^{*}\left(\left|
\vec{p}\right| \right)
A_{\vec{p}}^{L,a}\vec{B}_{\vec{p}}^a\right\} \exp \left\{
C_3\left(\left| \vec{p}\right| \right)
B_{\vec{p}}^{a+}A_{\vec{p}}^{L,a+}\right\} \mid 0\rangle \right]
^8  \nonumber \\ &&=\left[ \langle 0\mid \sum\limits_{m=0}^\infty
\left| C_3\left(\left| \vec{p}\right| \right) \right|
^{2m}\frac{\left(A_{\vec{p}}^{L,a}B_{\vec{p} }^a\right)
^m\left(B_{\vec{p}}^{a+}A_{\vec{p}}^{L,a+}\right)
^m}{\left(m!\right) ^2}\mid 0\rangle \right] ^8  \nonumber \\
&&=\left[ \sum\limits_{m=0}^\infty \left| C_3\left(\left|
\vec{p}\right| \right) \right| ^{2m}\right] ^8=\left[ \frac
1{\left(1-\left| C_3\left(\left| \vec{p}\right| \right) \right|
^2\right) }\right] ^8\qquad \text{for} \quad \left|
C_3\left(\left| \vec{p}\right| \right) \right| <1, \label{normLS}
\end{eqnarray}

\noindent in which we considered the relation

\begin{equation}
\langle 0\mid \left(A_{\vec{p}}^{L,a}B_{\vec{p}}^a\right)
^m\left(B_{\vec{ p }}^{a+}A_{\vec{p}}^{L,a+}\right) ^m\mid
0\rangle =\left(m!\right) ^2.
\end{equation}

\newpage

Finally the factor connected with the ghost fields can be
evaluated to be

\begin{eqnarray}
&&\left[ \langle 0\mid \left(1-iC_3^{*}\left(\left| \vec{p}\right|
\right) c_{\vec{p}}^a\overline{c}_{\vec{p}}^a\right)
\left(1+iC_3\left(\left| \vec{ p}\right| \right)
\overline{c}_{\vec{p}}^{a+}c_{\vec{p}}^{a+}\right) \mid 0\rangle
\right] ^8  \nonumber \\ &&=\left[ 1+\left| C_3\left(\left|
\vec{p}\right| \right) \right| ^2\langle 0\mid
c_{\vec{p}}^a\overline{c}_{\vec{p}}^a\overline{c}_{\vec{p}}^{a+}c_{
\vec{p}}^{a+}\mid 0\rangle \right] =\left[ 1-\left|
C_3\left(\left| \vec{p} \right| \right) \right| ^2\right] ^8.
\label{normG}
\end{eqnarray}

After substituting all the calculated factors, the norm of the
state can be written as

\begin{equation}
N=\langle \Phi \mid \Phi \rangle =\prod\limits_{\sigma =1,2}\left[
\sum\limits_{m=0}^\infty \left| C_\sigma \left(\left|
\vec{p}\right| \right) \right| ^{2m}\frac{\left(2m\right)
!}{\left(m!\right) ^2}\right] ^8.
\end{equation}

Therefore, it is possible to define the normalized state

\begin{equation}
\mid \widetilde{\Phi }\rangle =\frac 1{\sqrt{N}}\mid \Phi \rangle,
\end{equation}
\begin{equation}
\langle \widetilde{\Phi }\mid \widetilde{\Phi }\rangle =1.
\end{equation}

Note that, as should be expected, the norm is not dependent on the
$ C_3\left(\left| \vec{p}\right| \right) $ parameter which defines
the non-physical particle operators entering in the definition of
the state.

\section{Gluon and Ghost modified propagators.}

Let us determine the form of the main elements of perturbation
theory, that is the free particle propagators. It will be seen
that the propagators associated to the considered state has the
same form as proposed in [11] under a proper selection of the
parameters. Consider the generating functional of the free
particle Green's functions as given by

\begin{equation}
Z\left(J\right) \equiv \langle \widetilde{\Phi }\mid T\left(\exp
\left\{ i\int d^4x \sum \limits_{a=1,..,8} J^{\mu,a}\left(x\right)
A_\mu ^{a}\left(x\right) \right\} \right) \mid \widetilde{\Phi
}\rangle.
\end{equation}

As a consequence of the Wick's theorem the generating functional
can be written in the form \cite{Gasiorowicz}

{\setlength\arraycolsep{0.5pt}
\begin{eqnarray}
Z\left(J\right) &\equiv &\langle \widetilde{\Phi }\mid \exp
\left\{ i\int d^4x\sum \limits_{a=1,..,8}J^{\mu,a}\left(x\right)
A_\mu ^{a-}\left(x\right) \right\} \exp \left\{ i\int d^4x\sum
\limits_{a=1,..,8}J^{\mu,a}\left(x\right) A_\mu ^{a+} \left(x\right) \right\}
\mid \widetilde{\Phi }\rangle \nonumber \\
&&\times\exp \left\{ \frac i2\sum\limits_{a,b=1,..8}\int d^4xd^4y
J^{\mu ,a}\left(x\right) D_{\mu \nu
}^{ab}(x-y)J^{\nu,b}\left(y\right) \right\},
\end{eqnarray}
}where $D_{\mu \nu }^{ab}(x-y)$ is the usual gluon propagator.

Therefore, the sought for modification to the free propagator is
completely determined by the term

\begin{equation}
\prod\limits_{a=1,..,8}\langle \widetilde{\Phi }\mid \exp \left\{
i\int d^4xJ^{\mu,a}\left(x\right) A_\mu ^{a-}\left(x\right)
\right\} \exp \left\{ i\int d^4xJ^{\mu,a}\left(x\right) A_\mu
^{a+} \left(x\right) \right\} \mid \widetilde{\Phi }\rangle,
\label{Mod}
\end{equation}

\noindent where all the color dependent operators are decoupled
thanks to the commutation relations.

The annihilation and creation parts of the field operators in this
expression are given by

\begin{eqnarray}
A_\mu ^{a+}\left(x\right)
&=&\sum\limits_{\vec{k}}\left(\sum\limits_{\sigma
=1,2}A_{\vec{k},\sigma }^af_{k,\mu }^\sigma \left(x\right)
+A_{\vec{k}}^{L,a}f_{k,L,\mu }\left(x\right) +B_{\vec{k}
}^af_{k,S,\mu }\left(x\right) \right), \\ A_\mu
^{a-}\left(x\right)
&=&\sum\limits_{\vec{k}}\left(\sum\limits_{\sigma
=1,2}A_{\vec{k},\sigma }^{a+}f_{k,\mu }^{\sigma *}\left(x\right)
+A_{\vec{k}}^{L,a+}f_{k,L,\mu }^{*}\left(x\right) +B_{\vec{k}
}^{a+}f_{k,S,\mu }^{*}\left(x\right) \right).
\end{eqnarray}

For each color the following terms should be calculated

{\setlength\arraycolsep{0.5pt}
\begin{eqnarray}
&\exp &\left\{ i\int d^4xJ^{\mu,a}\left(x\right) A_\mu
^{a+}\left(x\right) \right\} \mid \Phi \rangle  \label{expd} \\
&=&\exp \left\{ i\int d^4xJ^{\mu,a}\left(x\right)
\sum\limits_{\vec{k} }\left(\sum\limits_{\sigma
=1,2}A_{\vec{k},\sigma }^af_{k,\mu }^\sigma \left(x\right)
+A_{\vec{k}}^{L,a}f_{k,L,\mu }\left(x\right) +B_{\vec{k}
}^af_{k,S,\mu }\left(x\right) \right) \right\}  \nonumber \\
&&\times \exp \left\{ C_1\left(\left| \vec{p}\right| \right)
A_{\vec{p} ,1}^{a+}A_{\vec{p},1}^{a+}+C_2\left(\left|
\vec{p}\right| \right) A_{\vec{p}
,2}^{a+}A_{\vec{p},2}^{a+}+C_3\left(\left| \vec{p}\right| \right)
\left(B_{ \vec{p}}^{a+}A_{\vec{p}}^{L,a+}+
i\overline{c}_{\vec{p}}^{a+}c_{\vec{p} }^{a+}\right) \right\} \mid
0\rangle.  \nonumber
\end{eqnarray}
}

After a systematic use of the commutation relations among the
annihilation and creation operators, the exponential operators can
be decomposed in products of exponential for each space-time mode.
This fact allows us to perform the calculation for each kind of
wave independently. Considering first the transverse components is
obtained

\begin{eqnarray}
&&\exp \left\{ i\int d^4xJ^{\mu,a}\left(x\right)
\sum\limits_{\vec{k}}A_{ \vec{k},\sigma }^af_{k,\mu }^\sigma
\left(x\right) \right\} \exp \left\{ C_\sigma \left(\left|
\vec{p}\right| \right) A_{\vec{p},\sigma }^{a+}A_{ \vec{p},\sigma
}^{a+}\right\} \mid 0\rangle \quad \text{for}\ \sigma =1,2
\nonumber
\\ &&=\exp \left\{ C_\sigma \left(\left| \vec{p}\right| \right)
\left(A_{\vec{ p},\sigma }^{a+}+i\int d^4xJ^{\mu,a}\left(x\right)
f_{p,\mu }^\sigma \left(x\right) \right) ^2\right\} \mid 0\rangle.
\label{52}
\end{eqnarray}

The expressions for the longitudinal and scalar modes can be
evaluated in a similar way and are found to be

\begin{eqnarray}
&&\exp \left\{ i\int d^4xJ^{\mu,a}\left(x\right)
\sum\limits_{\vec{k}}B_{ \vec{k}}^af_{k,S,\mu }\left(x\right)
\right\} \exp \left\{ i\int d^4xJ^{\mu,a}\left(x\right)
\sum\limits_{\vec{k}}A_{\vec{k} }^{L,a}f_{k,L,\mu }\left(x\right)
\right\}  \nonumber \\ &&\hspace{8.8cm}\times \exp \left\{
C_3\left(\left| \vec{p}\right| \right)
B_{\vec{p}}^{a+}A_{\vec{p}}^{L,a+}\right\} \mid 0\rangle \nonumber
\\ &&=\exp \left\{ C_3\left(\left| \vec{p}\right| \right)
\left(B_{\vec{p} }^{a+}-i\int d^4xJ^{\mu,a}\left(x\right) f_{p,L,\mu
}\left(x\right) \right) \right.  \label{53}
\\ &&\hspace{5.5cm}\left. \times \left(A_{\vec{p}}^{L,a+}-i\int
d^4xJ^{\mu,a}\left(x\right) f_{p,S,\mu }\left(x\right) \right)
\right\} \mid 0\rangle.  \nonumber
\end{eqnarray}

\noindent Here it should be noticed that the sign difference is
produced by the commutation relations.

For the calculation of the total modification (\ref{Mod}), it is
needed to evaluate

\begin{equation}
\langle \Phi \mid \exp \left\{ i\int d^4xJ^{\mu,a}\left(x\right)
A_\mu ^{a-}\left(x\right) \right\} =\left(\exp \left\{ -i\int
d^4xJ^{\mu,a}\left(x\right) A_\mu ^{a+}\left(x\right) \right\}
\mid \Phi \rangle \right) ^{\dagger }.  \label{iz}
\end{equation}

\noindent which may be easily obtained by the results for the
r.h.s of (\ref{52}) and (\ref{53}).

In what follows the following notation will be employed

\begin{equation}
J_{p,i}^a=\int \frac{d^4x}{\sqrt{2Vp_0}}J^{\mu,a}\left(x\right)
\epsilon _{i,\mu }\left(p\right).
\end{equation}

After using the relations (\ref{52}), (\ref{53}) and the result of
(\ref{iz}) the following factors are obtained for each color value
in the expression (\ref{Mod})

{\setlength\arraycolsep{0.1pt}
\begin{eqnarray}
&\langle 0\mid &\prod\limits_{\sigma =1,2}\exp \left\{ C_\sigma
^{*}\left(\left| \vec{p}\right| \right) \left(A_{\vec{p},\sigma
}^a-iJ_{p,\sigma }^a\right) ^2\right\} \exp \left\{ C_\sigma
\left(\left| \vec{p}\right| \right) \left(A_{\vec{p},\sigma
}^{a+}-iJ_{p,\sigma }^a\right) ^2\right\} \mid 0\rangle  \nonumber
\\ &\times &\langle 0\mid \exp \left\{ C_3^{*}\left(\left|
\vec{p}\right| \right) \left(A_{\vec{p}}^{L,a}+iJ_{p,S}^a\right)
\left(B_{\vec{p} }^a+iJ_{p,L}^a\right) \right\}  \nonumber \\
&&\qquad \times \exp \left\{ C_3\left(\left| \vec{p}\right|
\right) \left(B_{\vec{p}}^{a+}-iJ_{p,L}^a\right)
\left(A_{\vec{p}}^{L,a+}-iJ_{p,S}^a \right) \right\} \mid 0\rangle
\nonumber \\ &\times &\langle 0\mid \exp \left\{
C_3^{*}\left(\left| \vec{p}\right| \right)
\left(-ic_{\vec{p}}^a\overline{c}_{\vec{p}}^a\right) \right\} \exp
\left\{ C_3\left(\left| \vec{p}\right| \right)
\left(i\overline{c}_{\vec{p} }^{a+}c_{\vec{p}}^{a+}\right)
\right\} \mid 0\rangle.  \label{just3}
\end{eqnarray}
}

\noindent where the parts of the expression associated to each
space-time mode are also decoupled.

In evaluating these matrix elements the idea was the following.
First to expand the exponents of the exponential operators being
at the left of the scalar products in (\ref{just3}) by factorizing
the exponential operator having an exponent being linear in the
sources ``$J$''. After that, taking into account that the inverse
of this linear operator is annihilating the vacuum, it follows
that the net effect of this linear operator is to shift the
creation fields entering in the exponential operators at the right
of the scalar products in (\ref{just3}) in a constant being linear
in the sources ``$J$''. Further, the same procedure can be
performed to act with the exponential factor, which can be also
extracted from the new exponential operator acting on the vacuum
at the right. Now its action on the operators at the left of
(\ref{just3}) reduces again to a shift in a constant in the
annihilation fields defining this operator. In such a way it is
possible to arrive to a recurrence relation, which can be proven
by mathematical induction.

The recurrence relation obtained after n steps in the case of the
transverse modes takes the form

{\setlength\arraycolsep{0.1pt}
\begin{eqnarray}
&\exp &\left\{ -\left(J_{p,\sigma }^a\right) ^2\left[ C_\sigma
^{*}\left(\left| \vec{p}\right| \right) +4C_\sigma \left(\left|
\vec{p}\right| \right) \left(C_\sigma ^{*}\left(\left|
\vec{p}\right| \right) +\frac 12 \right) ^2\right. \right. \times
\nonumber \\ &&\hspace{4cm}\times \left. \left.
\sum\limits_{m=0}^n\left(4^{2m}\left(\left| C_\sigma \left(\left|
\vec{p}\right| \right) \right| ^2\right)
^{2m}+4^{2m+1}\left(\left| C_\sigma \left(\left| \vec{p}\right|
\right) \right| ^2\right) ^{2m+1}\right) \right] \right\}
\nonumber \\ &\times &\langle 0\mid \exp \left\{ C_\sigma
^{*}\left(\left| \vec{p} \right| \right) \left(A_{\vec{p},\sigma
}^a\right) ^2\right\} \times \nonumber \\ &&\qquad \times \exp
\left\{ -i2^{3+2n}J_{p,\sigma }^aA_{\vec{p},\sigma
}^a\left(C_\sigma ^{*}\left(\left| \vec{p}\right| \right) \right)
^{n+1}\left(C_\sigma \left(\left| \vec{p}\right| \right) \right)
^{n+1}\left(C_\sigma ^{*}\left(\left| \vec{p}\right| \right)
+\frac 12 \right) \right\}  \nonumber \\ &&\qquad \qquad \times
\exp \left\{ C_\sigma \left(\left| \vec{p}\right| \right)
\left(A_{\vec{p},\sigma }^{a+}\right) ^2\right\} \mid 0\rangle.
\label{65}
\end{eqnarray}
}

After restricting the possible values of $C_\sigma $ to satisfy
$\left| C_\sigma \left(\left| \vec{p}\right| \right) \right|
<\frac 12$ the linear part of the operators in the exponent is
multiplied by a quantity tending to zero in the limit $n=\infty $
and it can be omitted in such a limit. By also using the formula
for the geometrical series the following expression can be
obtained for (\ref{65})

{\setlength\arraycolsep{0.1pt}
\begin{eqnarray}
&\exp &\left\{ -\left(J_{p,\sigma }^a\right) ^2\left(C_\sigma
^{*}\left(\left| \vec{p}\right| \right) +4C_\sigma \left(\left|
\vec{p}\right| \right) \left(C_\sigma ^{*}\left(\left|
\vec{p}\right| \right) +\frac 12 \right) ^2\frac
1{\left(1-\left(2\left| C_\sigma \left(\left| \vec{p} \right|
\right) \right| \right) ^2\right) }\right) \right\}  \nonumber \\
&\times &\langle 0\mid \exp \left\{ C_\sigma ^{*}\left(\left|
\vec{p} \right| \right) \left(A_{\vec{p},\sigma }^a\right)
^2\right\} \exp \left\{ C_\sigma \left(\left| \vec{p}\right|
\right) \left(A_{\vec{p},\sigma }^{a+}\right) ^2\right\} \mid
0\rangle. \label{just4}
\end{eqnarray}
}

In a similar way for the factors in (\ref{just3}) corresponding to
longitudinal and scalar modes, it can be obtained

{\setlength\arraycolsep{0.5pt}
\begin{eqnarray}
&\exp &\left\{ -J_{p,S}^aJ_{p,L}^a\left(C_3^{*}\left(\left|
\vec{p}\right| \right) +C_3\left(\left| \vec{p}\right| \right)
\left(C_3^{*}\left(\left| \vec{p}\right| \right) +1\right) ^2\frac
1{\left(1-\left| C_3\left(\left| \vec{p}\right| \right) \right|
^2\right) }\right) \right\}  \nonumber \\ &\times &\langle 0\mid
\exp \left\{ C_3^{*}\left(\left| \vec{p}\right| \right)
A_{\vec{p}}^{L,a}B_{\vec{p}}^a\right\} \exp \left\{
C_3\left(\left| \vec{p}\right| \right)
B_{\vec{p}}^{a+}A_{\vec{p}}^{L,a+}\right\} \mid 0\rangle.
\end{eqnarray}
}

Therefore, after collecting the contributions of all the modes and
substituting $J_{p,i}^a$, by also assuming $2C_1\left(\left|
\vec{p}\right| \right) =2C_2\left(\left| \vec{p}\right| \right)
=C_3\left(\left| \vec{p} \right| \right) $ (which follows
necessarily in order to obtain Lorentz invariance) and using the
properties of the defined vectors basis, the modification to the
propagator becomes

\begin{eqnarray}
&&\exp \left\{ \frac 12\int
\frac{d^4xd^4y}{2p_0V}J^{\mu,a}\left(x\right)
J^{\nu,a}\left(y\right) g_{\mu \nu }\right.  \nonumber
\\ &&\hspace{2cm}\times \left. \left[ C_3^{*}\left(\left|
\vec{p}\right| \right) +C_3\left(\left| \vec{p}\right| \right)
\left(C_3^{*}\left(\left| \vec{p}\right| \right) +1\right) ^2\frac
1{\left(1-\left| C_3\left(\left| \vec{p}\right| \right) \right|
^2\right) }\right] \right\}.  \label{M0d2}
\end{eqnarray}

Now it is possible to perform the limit process
$\vec{p}\rightarrow 0$. In doing this limit, it is considered that
each component of the linear momentum $\vec{p}$ is related with
the quantization volume by

\[
p_x\sim \frac 1a,\ p_y\sim \frac 1b,\ p_z\sim \frac 1c,\ V=abc\sim
\frac 1{ \left| \vec{p}\right| ^3},
\]
Since $C_3\left(\left| \vec{p}\right| \right) <1$ then it follows

\begin{equation}
\lim_{\vec{p}\rightarrow 0}\frac{C_3^{*}\left(\left|
\vec{p}\right| \right) }{4p_0V}\sim \lim_{\vec{p}\rightarrow
0}\frac{C_3^{*}\left(\left| \vec{p} \right| \right) \left|
\vec{p}\right| ^3}{4p_0}=0.
\end{equation}

For the other limit it follows

\begin{equation}
\lim_{\vec{p}\rightarrow 0}\frac{C_3\left(\left| \vec{p}\right|
\right) \left(C_3^{*}\left(\left| \vec{p}\right| \right) +1\right)
^2\frac 1{ \left(1-\left| C_3\left(\left| \vec{p}\right| \right)
\right| ^2\right) } }{4p_0V},  \label{Lim1}
\end{equation}

Then, after fixing a dependence of the arbitrary constant $C_3$ of
the form $ \left| C_3\left(\left| \vec{p}\right| \right) \right|
\sim \left(1-\kappa \left| \vec{p}\right| ^2\right),\kappa >0$,
and $C_3\left(0\right) \neq -1$ the limit (\ref{Lim1}) becomes

\begin{equation}
\lim_{\vec{p}\rightarrow 0}\frac{C_3\left(\left| \vec{p}\right|
\right) \left(C_3^{*}\left(\left| \vec{p}\right| \right) +1\right)
^2\left| \vec{p} \right| ^3\frac 1{\left(1-\left(1-\kappa \left|
\vec{p}\right| ^2\right) ^2\right) }}{4p_0}=\frac C{2\left(2\pi
\right) ^4}  \label{const}
\end{equation}

\noindent where $C$ is an arbitrary constant determined by the
also arbitrary factor $\kappa$. An analysis of its properties has
been done which shows that $C$ can take only real and nonnegative
values.

Therefore, the total modification to the propagator including all
color values turns to be

{\setlength\arraycolsep{0.5pt}
\begin{eqnarray}
&\prod\limits_{a=1,..,8}&\langle \widetilde{\Phi }\mid \exp
\left\{ i\int d^4xJ^{\mu,a}\left(x\right) A_\mu
^{a-}\left(x\right) \right\} \exp \left\{ i\int
d^4xJ^{\mu,a}\left(x\right) A_\mu ^{a+}\left(x\right) \right\}
\mid \widetilde{\Phi }\rangle \nonumber \\ &=&\exp \left\{
\sum\limits_{a=1,..8}\int d^4xd^4yJ^{\mu,a}\left(x\right)
J^{\nu,a}\left(y\right) g_{\mu \nu }\frac C{2\left(2\pi \right)
^4} \right\}.
\end{eqnarray}
}

Also, the generating functional associated to the proposed initial
state can be written in the form

\begin{equation}
Z[J]=\exp \left\{ \frac i2\sum\limits_{a,b=1,..8}\int d^4xd^4y
J^{\mu ,a}\left(x\right) \widetilde{D}_{\mu \nu
}^{ab}(x-y)J^{\nu,b}\left(y\right) \right\},
\end{equation}

\noindent where

\begin{equation}
\widetilde{D}_{\mu \nu }^{ab}(x-y)=\int \frac{d^4k}{\left(2\pi
\right) ^4} \delta ^{ab}g_{\mu \nu }\left[ \frac 1{k^2}-iC\delta
\left(k\right) \right] \exp \left\{ -ik\left(x-y\right) \right\}
\label{propag}
\end{equation}

\noindent which shows that the gluon propagator has the same form
proposed in \cite{Cabo}, for the selected gauge parameter value
$\alpha =1$ (which corresponds to $\alpha=-1$ in that reference).

Finally, let us consider the possible modifications of the ghost
propagator, which can be produced by the new initial state. It is
needed to evaluate the expression

{\setlength\arraycolsep{0.5pt}
\begin{eqnarray}
\prod\limits_{a=1,..,8}\langle \widetilde{\Phi }\mid &\exp&
\left\{ i\int d^4x\left(\overline{\xi }^a\left(x\right)
c^{a-}\left(x\right) +\overline{ c}^{a-}\left(x\right) \xi
^a\left(x\right) \right) \right\}  \nonumber \\ \times &\exp&
\left\{ i\int d^4x\left(\overline{\xi }^a\left(x\right)
c^{a+}\left(x\right) +\overline{c}^{a+}\left(x\right) \xi
^a\left(x\right) \right) \right\} \mid \widetilde{\Phi }\rangle,
\label{ini}
\end{eqnarray}
}

In this case the calculation is easier because of the fermionic
character of the ghost makes that only two non-vanishing terms
exist in the series expansion of the exponential. Therefore, here
it is unnecessary to employ recurrence relations. The following
result for (\ref{ini}) is obtained
\begin{equation}
\exp \left\{ -\sum\limits_{a=1,..8}i\int d^4xd^4y\overline{\xi }^a
\left(x\right) \xi ^a\left(y\right) \frac{C_G}{\left(2\pi \right)
^4}\right\}
\end{equation}

\noindent where in this case $C_G$ is an arbitrary nonnegative
real constant. It will be equal to zero if taking
$C_3\left(0\right)$ real. This selection makes that the result
coincides with the one in the reference \cite{Cabo} where the
ghost propagator was not modified.

The expression of the generating functional for the ghost
particles takes the form

\begin{equation}
Z_G[\overline{\xi },\xi ]=\exp \left\{
i\sum\limits_{a,b=1,..8}\int d^4xd^4y \overline{\xi
}^a\left(x\right) \widetilde{D}_G^{ab}(x-y)\xi ^b\left(y\right)
\right\},
\end{equation}

\noindent where
\begin{equation}
\widetilde{D}_G^{ab}(x-y)=\int \frac{d^4k}{\left(2\pi \right) ^4}
\delta ^{ab}\left[ \frac{\left(-i\right) }{k^2}-C_G\delta
\left(k\right) \right] \exp \left\{ -ik\left(x-y\right) \right\}.
\end{equation}

Finally in order to illustrate one of the main properties of the
proposed modified perturbation expansion, let us review here a
previous calculation \cite{Cabo} of the gluon condensation
parameter $G^2$ in the ground state. In the simplest
approximation, that is the mean value of $G^2$ in the interaction
free initial state, it corresponds to evaluate

{\setlength\arraycolsep{0.5pt}
\begin{eqnarray}
\langle 0\mid S_{g}\left[ A\right] \mid 0\rangle &=&\left\{ \left[
\frac{1}{ 2i^2}S_{ij}^{g}\frac{\delta ^2}{\delta j_{i}\delta
j_{j}}+\frac{1}{3!i^3} S_{ijk}^{g} \frac{\delta ^{3}}{\delta
j_{i}\delta j_{j}\delta j_{k}}+\frac{1 }{4!i^4}
S_{ijkl}^{g}\frac{\delta ^{4}}{\delta j_{i}\delta j_{j}\delta
j_{k} \delta j_{l}}\right] \right\}  \nonumber \\ &&\times \exp
(\frac{i}{2}j_{i}\widetilde{D}_{ij}j_{j}),
\end{eqnarray}
}\noindent where, using the DeWitt notation, the symbol
$S_{ij...l}^g$ represents the functional derivative of the action
$S_{g}$ over a number of source arguments $j_{\mu
_i}^{a_i}(x_i),j_{\mu j}^{a_j}(x_j)$...and $j_{\mu
_l}^{a_l}(x_l)$. As usual in this convention, the equality of two
of compact indices $i,j...l$ means the sum over the color and
Lorentz indices and the subsequent integration over the spacetime
coordinates. The symbol $\widetilde{D}_{ij}$ is just the kernel of
the gluon propagator (\ref{propag}). The first and second terms in
the squared brackets have zero contribution as evaluated in
dimensional regularization at zero value of the sources. On the
other hand the last terms corresponding with four gluon
self-interaction gives a non vanishing addition to the gluon
condensation parameter precisely due to the condensate term in the
propagator.

The contribution can be evaluated to be

\[
\langle 0\mid S_g\left[ \phi \right] \mid 0\rangle \ =-\frac{72\
g^2C^2}{ (2\pi)^8}\int dx\,
\]

\noindent which corresponds with a gluon condensation parameter
given by
\[
G\ ^2\equiv \langle 0\mid G_{\mu \nu }^aG_{\mu \nu }^a\mid
0\rangle =\frac{ 288\ g^2C^2}{(2\pi)^8}.
\]

Therefore, it turns out that the procedure is able to predict the
gluon condensation at the most simple approximation.

\section{Summary}

By using the operational formulation of the Quantum Gauge Fields
Theory developed by Kugo and Ojima, a particular state vector of
QCD in the non-interacting limit, which obeys the BRST physical
state condition, was constructed. The general motivation for
looking for this wave-function is to search for a reasonably good
description of low energy QCD properties, through giving
foundation to the perturbative expansion proposed in \cite{Cabo}.
The high energy QCD description should not be affected by the
modified perturbative initial state. In addition it can be expect
that the adiabatic connection of the color interaction starting
with it as an initial condition, generate at the end the true QCD
interacting ground state. In case of having the above properties,
the analysis would allow to understand the real vacuum as a
superposition of infinite number of soft gluon pairs.

It has been checked that properly fixing the free parameters in
the constructed state, the perturbation expansion proposed in the
former work is reproduced for the special value $\alpha =1$ of the
gauge constant. Therefore, the appropriate gauge is determined for
which the expansion introduced previously \cite{Cabo} is produced
by an initial state, satisfying the physical state condition for
the BRST quantization procedure. The fact that the non-interacting
initial state is a physical one, lead to expect that the final
wave-function after the adiabatic connection of the interaction
will also satisfy the physical state condition for the interacting
theory. If this assumption is correct, the results of the
calculations of transition amplitudes and the values of physical
quantities should be also physically meaningful. In future, a
quantization procedure for arbitrary values of $\alpha $ will be
also considered. It is expected that with its help the gauge
parameter independence of the physical quantities could be
implemented. It seems possible that the results of this
generalization will lead to $\alpha $ dependent polarizations for
gluons and ghosts and their respective propagators, which however
could produce $\alpha $ independent results for the physical
quantities. However, this more involved discussion will be delayed
for future consideration.

It is important to mention now a result obtained during the
calculation of the modification to the gluon propagator, in the
chosen regularization. It is that the arbitrary constant $C$ was
determined here to be real and positive. This outcome restricts an
existing arbitrariness in the discussion given in the previous
work. As this quantity $C$ is also determining the square of the
generated gluon mass as positive or negative, real or imaginary,
therefore it seems very congruent to arrive to a definite
prediction of $C$ as real and positive.

The modification to the free ghost propagator introduced by the
considered vacuum state was also calculated. Moreover, after
considering the free parameter in the proposed trial state as
real, which it seems the most natural assumption, the ghost
propagator is not be modified, as it was assumed in \cite{Cabo}.

Some tasks which can be addressed in future works are: The study
of the applicability of the Gell-Mann and Low theorem with respect
to the adiabatic connection of the interactions, starting from the
here proposed initial state. The development of zero modes
quantization, that is gluon states with exact vanishing four
momentum. The ability to consider them with success would allow a
formally cleaner definition of the proposed state, by excluding
the auxiliary momentum $\vec{p}$ recursively used in the
construction carry out. Finally, the application of the proposed
perturbation theory in the study of some problems related with
confinement and the hadron structure.

\vspace{0.7cm}

{\bf Acknowledgments}

The authors would like to acknowledge the helpful comments and
suggestions of A. Gonzalez, F. Guzman, P. Fileviez, D. Bessis, G.
Japaridze, C. Handy A. Mueller, E. Weinberg and J. Lowenstein. One
of the authors (A.C.M.) is indebted by general support of the
Abdus Salam ICTP during stay (August to September 1999) in which
this work was prepared. The support of the Center of Theoretical
Studies of Physical Systems of the Clark Atlanta University and
the Christopher Reynolds Foundation, allowing the visit to U.S.A.
in which the results were commented with various colleagues, is
also greatly acknowledged.

\end{document}